\documentclass{article}

\usepackage{arxiv}

\usepackage[utf8]{inputenc} % allow utf-8 input
\usepackage[T1]{fontenc}    % use 8-bit T1 fonts
\usepackage{hyperref}       % hyperlinks
\usepackage{url}            % simple URL typesetting
\usepackage{booktabs}       % professional-quality tables
\usepackage{amsfonts}       % blackboard math symbols
\usepackage{nicefrac}       % compact symbols for 1/2, etc.
\usepackage{microtype}      % microtypography
\usepackage{lipsum}		% Can be removed after putting your text content
\usepackage{graphicx}

\title{The Threats of Artificial Intelligence Scale (TAI). Development,
Measurement and Test Over Three Application Domains.}
\author{ \href{https://orcid.org/0000-0002-6305-2997}{\includegraphics[scale=0.06]{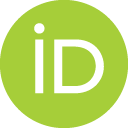}\hspace{1mm}Kimon~Kieslich} \\
	Department of Social Sciences\\
	University of Düsseldorf\\
	Universitätsstr. 1, Düsseldorf, 40225 \\
	\texttt{kimon.kieslich@hhu.de} \\
	\And
	\href{https://orcid.org/0000-0002-0553-7291}{\includegraphics[scale=0.06]{orcid.png}\hspace{1mm}Marco ~Lünich} \\
	Department of Social Sciences\\
	University of Düsseldorf\\
	Universitätsstr. 1, Düsseldorf, 40225 \\
	\texttt{marco.luenich@hhu.de} \\
	\AND
	Frank ~Marcinkowski \\
	Department of Social Sciences\\
	University of Düsseldorf\\
	Universitätsstr. 1, Düsseldorf, 40225 \\
	\texttt{frank.marcinkowski@hhu.de} \\
}

\renewcommand{\headeright}{}
\renewcommand{\undertitle}{}

\hypersetup{
pdftitle={The Threats of Artificial Intelligence Scale},
pdfsubject={q-bio.NC, q-bio.QM},
pdfauthor={Kimon~Kieslich, Marco ~Lünich, Frank ~Marcinkowski},
pdfkeywords={artificial intelligence, threat perceptions, fear, scale development},
}

\begin{document}
\maketitle
\sloppy
\raggedbottom

\begin{abstract}
  In recent years Artificial Intelligence (AI) has gained much popularity,
  with the scientific community as well as with the public. AI is often
  ascribed many positive impacts for different social domains such as
  medicine and the economy. On the other side, there is also growing
  concern about its precarious impact on society and individuals. Several
  opinion polls frequently query the public fear of autonomous robots and
  artificial intelligence (FARAI), a phenomenon coming also into scholarly
  focus. As potential threat perceptions arguably vary with regard to the
  reach and consequences of AI functionalities and the domain of
  application, research still lacks necessary precision of a respective
  measurement that allows for wide-spread research applicability. We
  propose a fine-grained scale to measure threat perceptions of AI that
  accounts for four functional classes of AI systems and is applicable to
  various domains of AI applications. Using a standardized questionnaire
  in a survey study (\emph{N}=891), we evaluate the scale over three
  distinct AI domains (loan origination, job recruitment and medical treatment).
 The data support the dimensional structure of the proposed
  Threats of AI (TAI) scale as well as the internal consistency and
  factoral validity of the indicators. Implications of the results and the
  empirical application of the scale are discussed in detail.
  Recommendations for further empirical use of the TAI scale are provided.
\end{abstract}

% keywords can be removed

\hypertarget{introduction}{%
\section{Introduction}\label{introduction}}

In recent years, applications making use of Artificial Intelligence (AI)
have gained re-newed popular interest. Expectations that AI might change
the face of various life domains for the better are abundant {[}7, 16,
36{]}. Be it medicine, mobility, scientific progress, the economy, or
politics; hopes are that AI will increase the veracity of input,
effectiveness and efficiency of procedures as well as the overall
quality of outcomes. Irrespective whether changes apply to the
workplace, public management, industries producing goods and services as
well as private life: As usual with the diffusion of new technologies
there is tremendous uncertainty as to how exactly developments will play
out {[}42{]}, what social consequences will manifest and to what extent
respective expectations of stakeholders and societal groups will
materialize. Oftentimes, there will be some people that immensely profit
from socio-technological innovations, while others are left behind and
cannot cope with the unfolding of events {[}13{]}. Thus, whenever new
technologies bring about social change, the success of their
implementation or failure depends upon the reaction of the affected
people. People might happily accept new technology, they might not care
nor use it at all, or they may even show severe reactance towards it
{[}3{]}. There is first empirical evidence suggesting that the general
public itself shows some considerable restraint when it comes to the
broad societal diffusion of AI applications or robots that might even
border on actual fear of such technology {[}22, 32, 37{]}. However, as
fear and respective threat perceptions are presuppositional theoretical
constructs, they necessitate a more fine-grained approach that goes
beyond broad claims of concerns or even fear regarding autonomous
systems.

Accordingly, in this paper, we argue for an improved assessment of the
perceived threats of AI and propose a survey scale to measure these
threat perceptions. First, a broadly usable measurement would need to
address perceived threats of AI as a precondition to any actual fear
experienced. This conceptual difference is subsequently based on the
literature on fear and fear appeals. Second, the perceived threat of AI
would need to take into account the context-dependency of respective
fears as most real-world applications of AI are highly domain-specific.
AI that assists in the medical treatment of a person's disease might be
perceived vastly different from an AI that takes over their job. Third,
not only do perceptions hinge on the domain in which people encounter AI
applications, it would also be necessary to differentiate between the
extent of an AI's actual autonomy and reach in inflicting consequences
upon a person. Thus, it needs to be asked to what extent the AI is
merely used for analysis of a given situation, or going even further,
whether the AI is used to actively give suggestions or even making
autonomous decisions.

As the field of application is crucial for the mechanism and effects of
threat perceptions concerning AI, any standardized survey measure needs
to be somewhat flexible and individually adaptable to accommodate the
necessities of a broad application that considers AI's functions and the
context of implementation. That is why our scale construction opts for a
design that can easily be adapted to varying research interests of AI
scholars.

Consequently, we developed a scale addressing threats of AI that takes
into account such necessary distinctions and subsequently tested the
proposed measure for three domains (i.e.~loan origination, job
recruitment, and medical treatment that are subject to an AI application)
in an online survey with German citizens. In our proposed measure of
perceived threats of AI, we aim to cover all aspects of AI functionality
and make it applicable to various societal fields, where AI applications
are used. Thereby, we highlight three contributions of our scale, that
are addressed in the following:

\begin{enumerate}
\def\labelenumi{\arabic{enumi})}
\item
  We underpin our scale development theoretically by connecting it with
  the psychological literature on fear appeals.
\item
  The construction of the scale differentiates between the discrete
  functionalities of AI that may cause different emotional reactions.
\item
  Moreover, we consider perceived threats of AI as dependent on the
  context of the AI's implementation. This means that any measure must
  pay respect to AI's domain-specificity.
\end{enumerate}

The collected data supports the factorial structure of the proposed TAI
scale. Furthermore, results show that people differentiate between
distinct AI functionalities, in that, the extent of the functional reach
and autonomy of an AI application evoke different degrees of threat
perceptions irrespective of domain. Still, such distinct perceptions do
also differ between the domains tested. For instance, recognition and
prediction with regard to a physical ailment as well as the
recommendation for a specific therapy made by an AI do not evoke
substantial threat perceptions. Contrarily, autonomous decision-making
in which an AI unilaterally decides on the proscribed treatment was met
with relatively bigger apprehension. At the same time, the application
of AI in medical treatment was generally perceived as less fearsome than
situations where AI applications are used to screen applicants on a job
or a financial loan.

Eventually, to measure construct validity, we assessed the effects of
the \emph{Threats of Artificial Intelligence} (TAI) scale on emotional
fear. Threat perceptions are a necessary, but not sufficient
prerequisite to fear. While most research directly focuses on fear, we
will subsequently argue for the benefits of addressing the preceding
threat perceptions. Ultimately, the threat perceptions do in fact
trigger emotional fear. Lastly, we discuss the adoption and use of the
TAI scale in survey questionnaires and make suggestions for its
application in empirical research as well as general managerial
recommendations with regard to public concerns of AI.

\hypertarget{public-perceptions-of-recent-developments-in-artificial-intelligence}{%
\section{Public perceptions of recent developments in Artificial
Intelligence}\label{public-perceptions-of-recent-developments-in-artificial-intelligence}}

In recent years there has been a somewhat re-newed interest in
applications of AI based on recent developments in computer technology
that allows for use of extensive processing power and the analysis of
vast amounts of so-called Big Data applications of Machine Learning,
Deep Learning and Neural Networks. Such applications gather under the
label of AI, which is ascribed a huge impact on society as a whole
{[}29{]}. Thereby, AI has especially seen widespread use in business and
public management {[}58{]}. As a consequence, the public discourse
regarding AI is mainly driven by companies that provide AI technology
looking for customers and markets for their products {[}6, 7, 28{]}.
Meanwhile, empirical evidence from survey research supports the
assumption that AI is not per se perceived as entirely positive by the
public. A cross-national survey by Kelley et al.~{[}29{]} shows that AI
is connected with positive expectations in the field of medicine, but
reservations are prevalent concerning data privacy and job loss. Another
concern is raised by Araujo et al.~{[}2{]}, who state that citizens
perceive high risks regarding decision-making AI. Moreover, a
representative opinion poll by Zhang and Dafoe {[}60{]} illustrates that
Americans as well as citizens from the European Union (EU) believe that
robots and AI could have harmful consequences for societies and should
be carefully managed. Additionally, Gnambs and Appel {[}19{]} show that
attitudes towards robots have recently changed for the worse in the EU.
Especially, when it comes to the influence of robots in the economy and
the substitution of workforce, people express fear {[}22, 37{]}. On a
broader level, a recent study by {[}32{]} inquired about the fear of AI
even found that a considerable amount of all Americans reported fears
when it comes to autonomous robots and AI.

\hypertarget{measuring-the-fear-of-autonomous-robots-and-artificial-intelligence}{%
\section{Measuring the Fear of Autonomous Robots and Artificial
Intelligence}\label{measuring-the-fear-of-autonomous-robots-and-artificial-intelligence}}

Using data from the Chapman Survey of American Fears, Liang and Lee
{[}32{]} set out to investigate the prevalence of \emph{fear of
autonomous robots and artificial intelligence} (FARAI). They come to the
conclusion that roughly a quarter of the US population experienced a
heightened level of FARAI. In the respective study, participants were
confronted with the question ``How afraid are you of the following?''.
The FARAI-scale was afterwards built out of these four items: 1)
``Robots that can make their own decisions and take their own
actions,'', 2) ``Robots replacing people in the work-force'', 3)
``Artificial intelligence'' and 4) ``People trusting artificial
intelligence to do work''. All items were rated on a four-point Likert
scale with answers ranging from ``not afraid (1), slightly afraid (2),
afraid (3), to very afraid (4)'' {[}32{]}.

While the authors shed some first light in addressing threat perceptions
of artificial intelligence and generate valuable insights into various
associations of FARAI with demographic and personal characteristics,
there is also need for a potential enhancement of the existing
measurement of FARAI. As the FARAI scale was developed out of a broad
questionnaire concerning many possible fears people in the US might
have, the measurement was not specifically developed for measuring the
distinctive fear of robots and AI, respectively. The FARAI scale also
varies in its scope. While item 3 broadly queries the fear of AI in
general, item 2 specifically inquiries about its specific impacts on the
economic sector. Items 1 and 4 query a specific functionality of AI,
with item 4 focusing on the human-machine connection. Thus, the items
are mixed in their expressiveness and aim at different aspects of AI's
impact. Accordingly, the scale does not allow for distinct assessments
of AI and necessary specifications concerning its domain of application
and the employed functions.

Besides, the public understanding of robots and AI might be influenced
by popular imaginations from pop-culture, science-fiction, and the
media, as is also already implied by Liang and Lee {[}32{]} and Laakasuo
et al.~{[}30{]}. Due to the popularity of vastly different types of
autonomous robots and AI in literature, film and comics, it is hard to
pin down what exactly comes to a person's mind inquired about both
terms. Delineating boundaries may not be possible when it comes to the
public imagination. As a survey research in the UK by Cave et
al.~{[}8{]} suggests, a quarter of the British population conflates the
term AI with robots. Accordingly, a conceptual clarification concerning
the distinction between the terms \emph{robot} and \emph{artificial
intelligence} is required to begin with. In the FARAI measure by Liang
and Lee {[}32{]}, there is a mixture between both terms as two question
items focus on each term, respectively. This terminological distinction
is often conflated in empirical research {[}14{]}. We believe that the
mixture of the terms might lead to avoidable ambiguity and maybe even
confusion, since people may think of two distinct and even completely
different phenomena or might not be able to distinguish between the two
constructs at all. According to the Oxford English Dictionary a robot is
``an intelligent artificial being typically made of metal and resembling
in some way a human or other animal'' or ``a machine capable of
automatically carrying out a complex series of movements, esp.~one which
is programmable'' while AI is ``the capacity of computers or other
machines to exhibit or simulate intelligent behaviour''. There is
certainly some conceptual overlap by definition, especially with regard
to the capacity of intelligent behavior demonstrated by an artificial
construct, hence, something that does not exist naturally, but is
human-made. It also cannot be ruled out that appraisal of robots may be
strongly associated with AI, especially when such robots are depicted as
autonomous.

Recently, the term AI, particularly, has renewedly received widespread
attention and describes techniques from computer science that gather
many different concepts like \emph{machine learning}, \emph{deep
learning} or \emph{neural networks}, which are the basis of autonomous
functionality and pervasive implementation. As a consequence, we decided
to focus our measurement solely on AI as it depicts the core issue of
the nascent technology, i.e.~autonomous intelligent behavior, which
applies to many use cases that do not necessarily include a physical
machine in motion.

\hypertarget{threat-perceptions-as-precondition-of-fear}{%
\section{Threat Perceptions as Precondition of
Fear}\label{threat-perceptions-as-precondition-of-fear}}

There is plenty of literature on the subject of fear, especially from
the field of human psychology. Altogether, fear is defined as a negative
emotion that is aroused in response to a perceived threat {[}40{]}. When
it comes to the origins of emotion, many studies rely on the appraisal
theory of emotion. Smith and Lazarus argue that``[t]he appraisal task for the person is to evaluate
perceived circumstances in terms of a relatively small number of
categories of adaptional significance, corresponding to different types
of benefit or harm, each with different implications for coping''
{[}51: 616{]}. Accordingly, the authors define relational themes for
different emotions. According to Smith and Lazarus {[}51{]} anxiety, respectively fear, is
evoked, when people perceive an ambiguous danger or threat, which is
motivationally relevant as well as incongruent to their goals. Thereby,
a threat is seen as an ``environmental characteristic that represents
something that portends negative consequences for the individuum''
{[}38: 3{]}. Furthermore, people perceive low or uncertain coping
potential. In other words: Fear is the result of a persons' appraisal
process, where a situation or an object is perceived as threatening and
relevant as well as no avoiding potential can be seen. If theses
appraisals are processed, people react with fear and try to avoid the
threat {[}51{]}, i.e.~in turning away from the object.

Many scholars build on appraisal theory to develop more specified
theories on the mechanisms of fear. Especially in health communication
much work on so called fear appeal literature has been done {[}38,
39{]}. In a nutshell, most fear appeal theories state that a specific
object, event or situation (e.g., a disease) threatens the well-being of
a person {[}46, 59{]}. With the development of the Extended Parallel
Process Model (EPPM), Witte {[}59{]} theorizes that this threat is at
first processed cognitively. Thereby, severity and susceptibility of the
threat as well as coping potential (self and general), i.e.~the amount
of efficacy {[}59{]} respectively control {[}10{]}, is rated. Depending
on the weights of these cognitive apprehensions, people react
differently. Fear emerges, when the threat perception is high, while the
coping perception is low. As a result, message denial arises that is
mostly characterized by avoiding the threat. On the other hand, when
threat as well as coping potential are perceived as high, message
acceptance results. If this happens, people actively engage with the
threat, for example in gathering information about the threat or
actively combating potential harms. In this case, fear does not emerge.
Whereas the empirical examination of the EPPM found no clear proof
{[}41{]} and many suggestions for extending the model have been made
{[}52, 53{]}, scholars agree upon the central persuasion effects of
threat and coping perceptions {[}50{]}. Moreover, the EPPM commonly
serves as a framework for further research {[}33{]}.

Transferred to the subject of perceived threats of AI, we belief that AI
is best described as an environmental factor that might cause fear.
However, AI should not be treated as a specific fear itself. In our
view, fear may be a result of a cognitive appraisal process, where AI
depicts a potential threat origin. Thus, we explicitly focus on threats
of AI, not fear of AI. This idea becomes more prevalent in thinking
about an actual situation. For example, a person is confronted with an
AI system that decides over an approval of a credit. This person most
likely will not be afraid of the computer system, but will rather
evaluate cognitively the threat that such a system might pose to its
well-being. The person then rates the probability of the system causing
harm (e.g., if it denies the credit). If the outcome of this process
ends in a negative evaluation for the person, fear will be evoked.
However, this fear is based on the threat the AI systems poses and not
on the AI system itself. This is crucial for our understanding of
threats of AI.

\hypertarget{context-specificity-of-threat-perceptions}{%
\section{Context Specificity of Threat
Perceptions}\label{context-specificity-of-threat-perceptions}}

It is also important to address the social situation, in which a threat
is perceived. Smith and Lazarus {[}51{]} already stated that an ``appraisal can, of
course, change (1) as the person-environment relationship changes; (2)
in consequence of self-protective coping activity (e.g.~emotion-focused
coping); (3) in consequence of changing social structures and culturally
based values and meanings; or (4) when personality changes, as when
goals or beliefs are abandoned as unservicable'' (624). Furthermore, Tudor
{[}56{]} proposed a sociological approach for the understanding of fear.
He developed a concept, in which he distinguishes parameters of fear
including environments, cultures as well as social structures.

Thereby, contexts can vary in manifold ways. A rather simple example for
what Tudor {[}56{]} refers to as an environmental context is the case of the
wild animal: for instance, a tiger could face a human being; however,
arguably there is a huge difference in fear reaction if one is
confronted with the tiger in a zoo or in its natural habitat. Thus, the
environmental factor ``cage'' does have a huge impact on the incitement
of fear. Additionally, cultural backgrounds can affect the way threats
are perceived: ``If our cultures repeatedly warn us that this kind of
activity is dangerous, or that sort of situation is likely to lead to
trouble, then this provides the soil in which fearfulness may grow''
{[}56: 249{]}. Lastly, social structures described that the societal role of
an individual might influence threat perceptions. For instance, that
could be the job position of an employee or just the belonging to a
specific societal group.

Furthermore, different social actors are able to influence the social
construction of public fears, i.e.~if and how environmental stimuli are
treated as threats {[}10{]}. According to Dehne {[}10{]}, the creation
of fear, among other factors, is dependent upon transmission of
information in a society. In that, especially scientific, economic,
political and media actors affect the social construction of threats.
However, depending on the actors that take the highest share in the
public discourse, different threat perceptions might emerge. For
instance, given an AI application in medicine, we assume that science as
well as media lead the debate. On the other hand, an AI application in
the field of recruiting will probably be led by economic actors. It is
plausible that there are specific context dependencies (who informs the
public about a specific AI application) that have an influence on
(threat) perceptions.

In summary, there are many (social) factors that shape the way emotions
are elicited leading to the conclusion that threat perceptions heavily
rely on the context in which an individual encounters AI. Of course, we
are not able to cover all possible contexts of AI related threats.
However, we distinguish two context groups, which are important for the
understanding of TAI: AI functionality and distinct domains of AI
applications.

\hypertarget{distinct-dimensions-of-ai-functionality}{%
\subsection{Distinct Dimensions of AI
Functionality}\label{distinct-dimensions-of-ai-functionality}}

What an AI is capable or supposed to do may have a decisive effect on
the appraisal of AI applications. However, AI is a generic term that
unites many different functionalities. In the scientific community,
there are manifold definitions on the term AI and what can and what
cannot be counted as an AI system. Whereas there is not one definition,
most scholars agree upon central functionalities AI systems can perform.
Nevertheless, there is no consensus upon how to group these
functionalities. For example, Hofmann {[}24{]}, identify perceiving,
identification, reasoning, predicting, decision-making, generating and
acting as AI functions. Though, we base our approach on the periodic
systems of AI {[}4{]} and group AI functionalities into four categories,
which undoubtedly intersect each other: recognition, prediction,
recommendation and decision-making.

Noteworthy, our approach is quite similar to Hofmann {[}24{]}: however,
we subsumed generating and acting into the category of decision-making
as we focus on AI that act autonomously in that category. Additionally,
we added perceiving and identification into one category: This decision
was based upon the results of a pre-test of the scale, which we
conducted with 304 participants. Our results show that participants
could not differentiate between the perceiving and identification
function.

In the following, we elaborate on our proposed AI function classes:

\hypertarget{recognition}{%
\subsubsection{Recognition}\label{recognition}}

Recognition describes the task of analyzing input data in various forms
(e.g., images, audio) and recognizing specific patterns in the data.
Depending on the application, these patterns can vary hugely. In a
health application, AI recognition is used to detect and identify breast
cancer {[}45{]}. In the economic sector AI systems promise to detect
(personal) characteristics and abilities of potential employees via
their voices and / or faces {[}23{]}.

\hypertarget{prediction}{%
\subsubsection{Prediction}\label{prediction}}

In prediction tasks, AI applications prognose future conditions on the
basis of the analyzed data. It differentiates from recognition in
forecasting developments of specific states, whereas recognition mostly
classifies the given data. In the medical sector, a AI applications are
able to calculate the further development of diseases on basis of
medical diagnoses and (statistical) reports {[}9{]}.

\hypertarget{recommendation}{%
\subsubsection{Recommendation}\label{recommendation}}

Recommendation describes a task in the field of human-computer
interaction. Thereby, AI systems directly engage with humans, mostly
decision makers, in recommending specific actions. These actions are
again highly dependent on the actual application field. For the medical
example this could mean that the AI application, which takes into
account all given data, proposes a medical treatment to the doctor
{[}11{]}. Noteworthy, the decision to accept or decline this suggestion
is still made by the physician or the patient, respectively.

\hypertarget{decision-making}{%
\subsubsection{Decision-making}\label{decision-making}}

Ultimately, the functionality decision-making refers to AI systems that
operate autonomously. Oftentimes, these applications are also called
algorithmic decision-making (ADM) systems. Hereby, AI systems learn and
act autonomously after being carefully trained by developers. The most
prominent application is with no doubt autonomous driving {[}35{]}.
However, decision-making tasks can also be found in other domains of
application. For example, in medicine AI systems could directly decide
over medical treatments of patients; in the higher education sector ADM
could decide about the admissions of students' applications to
university {[}34{]}. Concerning the human-computer interaction an ADM
substitutes the human task completely.

Noteworthy, two points are important to mention. Firstly, the
functionalities can depend on each other and are thus not completely
separable. Secondly, AI applications in specific fields do not
necessarily have to fulfill all AI functionalities. Mostly, AI systems
just perform one task while not providing the other ones. We stress that
threat perceptions of AI - in technical terms - should not be treated as
a second-order factor. Rather our scale deploys a toolbox which can be
used to cover threat perceptions of the different functionalities.
However, we expect that there are significant correlations between the
functionalities. In conclusion, we pose the following research question:

\emph{RQ1: Do respondents have distinct threat perceptions regarding the
different functions AI systems perform?}

\hypertarget{distinct-domains-of-ai-application}{%
\subsection{Distinct Domains of AI
Application}\label{distinct-domains-of-ai-application}}

As usual, social science research addresses the social change induced by
technological phenomena and artifacts in various domains of public and
private life. Depending on the domain of application, AI may be
wholeheartedly welcomed or seen as a severe threat {[}14, 15{]}. For
instance, imagining to hand over certain decisions to an AI may appear
rather innocuous for certain lifestyle choices such as buying a product
or taking a faster route to a destination, but may lead to reactance
when perceived individual stakes are high, e.g.~when AI interferes in
life altering decisions that affect one's health or career decisions. As
applications of AI are expected to get implemented in manifold life
domains, research will need to address the respective perceptions of the
people affected. The domain specificity of effects is already an
established approach in social science research; for instance Acquisti et al. {[}1{]} as
well as Bol et al. {[}5{]} found that distinct application domains do matter in
terms of online privacy behavior. Additionally, Araujo et al. {[}2{]} analyzed
perceptions of automated decision-making AI in three distinct domains.
Thus, we believe that a measurement of threat perceptions also needs to
be adaptable to a multi-faceted universe of AI related phenomena, some
of which might not even be known to date. Concludingly, we propose a
measurement that is adaptable to every AI domain. As follows, the
proposed TAI scale is tested in three different domains, namely loan
origination, job recruitment, and medical treatment.

\hypertarget{loan-origination---assessing-creditworthiness}{%
\subsubsection{Loan Origination - Assessing
creditworthiness}\label{loan-origination---assessing-creditworthiness}}

AI technologies are already applied in the finance sector, i.e.~in
credit approval {[}18{]}. As credit approval is a more or less
mathematical problem, it is reasonable that AI based algorithms are
applied for this purpose. The algorithms used analyze customer data and
calculate potential payment defaults - and finally can decide, whether a
credit is approved. As individual goals greatly depend on such
decisions, it may pose a threat for individuals, who believe that their
input data might be deficient or assume that the processing biased.

\hypertarget{job-recruitment---assessing-the-qualification-and-aptitude-of-applicants}{%
\subsubsection{Job Recruitment - Assessing the qualification and
aptitude of
applicants}\label{job-recruitment---assessing-the-qualification-and-aptitude-of-applicants}}

Recently, AI applications have been applied to the field of human
resource management, i.e.~recruiting {[}48{]}. More specifically, AI can
be used to analyze and predict performance of employees. Furthermore, AI
based systems are able to recommend or select potential job candidates
{[}54{]}. However, there are several potential risks of the use of AI
systems in human resource management. For instance, algorithms based on
existing job performance data may be biased and lead to discrimination
of specific population groups {[}43, 48{]}.

\hypertarget{health---medical-treatment-of-diseases}{%
\subsubsection{Health - Medical Treatment of
Diseases}\label{health---medical-treatment-of-diseases}}

One of the most important fields of AI development and implementation is
with no doubt health care/medicine {[}26{]}. Especially, in fields where
imaging techniques are applied (e.g., radiology) AI applications are
frequently used {[}55{]}. Recent works show that AI applications are
especially appropriate to detect and classify specific diseases, for
example breast or skin cancer, in X-ray images {[}26, 55{]}. Moreover,
another AI application can identify gene-related diseases in face images
of patients {[}21{]}. Generally, people tend to have optimistic
perceptions of the use of AI in medicine {[}29{]}.

Summing up, it may be assumed that distinct domains of AI application
cause different threat perceptions. As mentioned earlier, a possible
explanatory approach is that the public discourse, through which
individuals are mostly confronted with AI, is led by different actor
groups. Another reason to believe that domains do vary is the actual
tasks AI systems perform and which severity individuals ascribe to them.
Presumably, also personal relevance appraisals play a major role in the
level of threat individuals ascribe to distinct domains. An individual,
who does not plan to apply for a credit will probably rate the use of an
AI system for credit approval as less threatening than a person who is
in dire need of a loan. Arguably, we can only focus on a small sample of
potential AI domains. In conclusion, we formulate the following
hypothesis:

\emph{H1: Threat perceptions of AI differ between distinct domains of AI
application.}

\hypertarget{fear-reactions-towards-ai}{%
\section{Fear Reactions Towards AI}\label{fear-reactions-towards-ai}}

As outlined in section four, perceived threats of AI are a precondition
of emotional (fear) reactions. Thus, we assume that threat perceptions
concerning AI actually trigger fear reactions. Accordingly, hypothesis 2
reads as followed.

\emph{H2: Threat perceptions of AI induce fear among respondents.}

As threat perceptions of AI functionalities and domains may differ
vastly from each other, we are interested in whether the amount of
perceived fear (if any) that is explained by our proposed measure also
differs by context. Arguably, not all threat perceptions necessarily
need to cause the same fear reactions. For instance, if subjects
perceive high levels of efficacy in dealing with the potential threat, a
far less strong emotional reaction is likely to occur {[}59{]}. This
becomes particularly obvious, when comparing the recommendation and the
decision-making functionality. Decision-making AI takes control away
from the individual, whereas in recommendation at least an (other) human
still has control over the process.

Therefore, we test whether the measure is able to capture induced fear
reactions across the different contexts and whether these differ.
Accordingly, we pose and address the following research question:

\emph{RQ2: Does the influence of threat perceptions of AI on fear differ
between contexts?}

\hypertarget{method}{%
\section{Method}\label{method}}

Accordingly, we set out to develop a measurement scale for the
application in survey research on AI that addresses the threat
perceptions of people that are confronted with various forms of AI
implementation.

The aim is to reliably and validly assess the extent to which
respondents perceive autonomous systems as a threat to themselves.
Moreover, the scale needs to be standardized allowing for comparisons
between samples from various populations, but flexible enough allowing
for application in distinct domains of AI research. It thus needs to be
as concise as possible affording to be included in brief questionnaires.

We tested our scale using a non-representative German online access
panel. We used an online questionnaire with a split survey design
comparing the threat perceptions towards AI in the three sectors loan
origination, job recruitment, and medical treatment. Participants were
randomly assigned to one of the three groups. We chose the method of
matching urns after completion of the questionnaire to ensure an even
distribution among the different questionnaire groups.

\hypertarget{sample}{%
\section{Sample}\label{sample}}

Participants were recruited from the SoSci Open Access Panel between the
30th September and 14th October 2019 {[}31{]}. All in all, 917 subjects
completed the questionnaire. In the data cleaning process, we had to
drop 26 cases from our data set. Data elimination was based on two
criteria: minus scores deployed by the access panel as well as time for
completing the questionnaire. The minus scores are calculated on basis
of the sum of the deviations from the average time for answering
individual questionnaire pages in order to identify respondents'
inattentiveness. For our data cleaning process, we chose the access
panel's recommended value for a conservative cut-off criterion to ensure
a high-quality data set. Moreover, we checked the overall time score
participants needed to fill out the questionnaire. We excluded all data
from participants with an answering time below five minutes, since we
defined this minimum value after a pre-test of our questionnaire. Thus,
our final sample consists of \emph{n}=891 participants.

Turning to the distribution by questionnaire groups, 296 participants
were assigned to the `loan origination' group, 294 to the `job
recruitment' group and 301 to the `medical treatment' group. Of all
participants 445 identify as female (50.0\%) and 438 as male (49.2\%),
while seven (0.8\%) respondents identify as non-binary. The average age
of the participants is approximately 46 years (SD=15.66). Because of the
demographic structure of the access panel, 82 percent of our
participants have the highest German school leaving certificate.
Arguably, our data is not representative for the German population,
which should be acknowledged when interpreting the descriptive results.
However, our primary interest of introducing a valid scale remain
unchallenged by this limitation.

\hypertarget{measurement}{%
\section{Measurement}\label{measurement}}

\hypertarget{threat-perceptions-of-artificial-intelligence}{%
\subsection{Threat Perceptions of Artificial
Intelligence}\label{threat-perceptions-of-artificial-intelligence}}

We propose a measurement for threat perceptions concerning AI based on
the specific functionality that AI systems can perform. We identified
`recognition', `prediction', `recommendation' and `decision-making' as
the core functions current of AI systems performance from a user's
perspective. The phenomenon that is AI was firstly explained to the
participants with a short text, which also contained the information
that AI currently draws widespread public attention. Furthermore, a
broad definition of AI systems and functionality was given in a neutral
tone as well as an explanation of how AI systems could be used in the
specific context presented to the respondents.

To achieve context independence, we then developed formal items with
clozes for the specific thematic foci, which can be seen as a toolbox
that is customizable for distinct areas of application. Altogether,
participants had to rate twelve statements on 5-point Likert scales
(1=``non-threatening'' to 5=``very threatening''). The question block
with the statements was introduced with the following text: ``If you now
think of the use of AI in {[}specific context{]}, how threatening do you
think computer applications of artificial intelligence are
that\ldots{}''. The items for the specific functionalities reads as
follows:

Recognition: ``(\ldots) detect (object)'' (RCG1), ``(\ldots) record
(object)'' (RCG2) and ``(\ldots) identify (object)'' (RCG3).

Prediction: ``(\ldots) forecast the development of (object)'' (PDC1),
``(\ldots) predict the development of (object)'' (PDC2) and ``(\ldots)
calculate the development of (object)'' (PDC3).

Recommendation: ``(\ldots) recommend (action)'' (RCM1), ``(\ldots)
propose (action)'' (RCM2) and ``(\ldots) suggest (action)'' (RCM3).

Decision-making: ``(\ldots) decide on (action)'' (DSM1), ``(\ldots)
define (action)'' (DSM2) and ``(\ldots) preset (action)'' (DSM3).

The brackets for \emph{object} were filled with the terms I)
``diseases'', II) ``suitability of applicants/work performance'', and
III) ``probability of default of credits/creditworthiness''. The
brackets for \emph{action} were filled with the terms I) ``medical
treatment'', II) ``hiring applicants'', and III) ``granting of
credits''. An example sentence reads as follows: ``(\ldots) how
threatening do you think computer applications of artificial
intelligence are that recommend a medical treatment.''

\hypertarget{emotional-fear-reaction}{%
\subsection{Emotional Fear Reaction}\label{emotional-fear-reaction}}

Lastly, emotional responses towards the use of AI in the respective
domain of application were retrieved. Participants had to rate how
strongly they experience the emotion of fear on 5-point Likert scales
(1=``non at all'' to 5=``very strong''). Fear was measured through the
items ``afraid'' (FEAR1), ``frightened'' (FEAR2), and ``anxious''
(FEAR3) {[}44{]}.

\hypertarget{results}{%
\section{Results}\label{results}}

To test our measurement, we performed confirmatory factor analyses (CFA)
with the \emph{lavaan} package {[}47{]} in R (version 4.0) as well as
several test statistics with the \emph{semTools package} {[}27{]}. For
visualization we used the \emph{semPlot} package {[}12{]}. Firstly, we
calculated a CFA with configural invariance. To check, if the factor
loadings differ between the applications, we secondly calculated a model
with measurement invariance. Thirdly, we constrained the intercept of
the measurement to check for scalar invariance, i.e.~to analyze whether
threat perceptions are different between application areas. Lastly, we
set one intercept free to gain our final model; that is, our results
suggest the TAI scale as a measurement with partial scalar invariance.
To check the influence of the TAI scale on emotional fear, we built a
structural equation model (SEM) with emotional fear as dependent
variable. We will further elaborate on our findings.

\hypertarget{descriptives}{%
\subsection{Descriptives}\label{descriptives}}

We first calculated descriptive statistics (mean, standard deviation,
skewness and kurtosis) for all scale items separately for each domain
(table 1). The descriptive values for each threat perception of the
distinct AI functionalities are quite equal between the domains but
differ considerably between different functionalities of AI. For
example, we see that the decision-making functionality provoked the
highest threat perceptions in all domains.

\begin{table*}

\caption{\label{tab:unnamed-chunk-4}Descriptives}
\centering
\resizebox{\linewidth}{!}{
\fontsize{9}{11}\selectfont
\begin{tabular}[t]{lrrrrrrrrrrrr}
\toprule
\multicolumn{1}{c}{ } & \multicolumn{4}{c}{Loan Origination} & \multicolumn{4}{c}{Job Recruitment} & \multicolumn{4}{c}{Medical Treatment} \\
\cmidrule(l{3pt}r{3pt}){2-5} \cmidrule(l{3pt}r{3pt}){6-9} \cmidrule(l{3pt}r{3pt}){10-13}
Item & M & SD & Skewness & Kurtosis & M & SD & Skewness & Kurtosis & M & SD & Skewness & Kurtosis\\
\midrule
RCG1 & 3.041 & 1.229 & -0.066 & -1.004 & 3.173 & 1.133 & -0.075 & -0.740 & 1.841 & 0.974 & 1.185 & 0.914\\
RCG2 & 3.057 & 1.230 & -0.022 & -0.965 & 3.088 & 1.188 & 0.000 & -0.931 & 2.033 & 1.146 & 0.930 & -0.107\\
RCG3 & 3.162 & 1.233 & -0.103 & -0.960 & 3.235 & 1.204 & -0.175 & -0.980 & 1.887 & 1.020 & 1.165 & 0.695\\
PDC1 & 2.865 & 1.214 & 0.009 & -1.003 & 3.765 & 1.140 & -0.595 & -0.602 & 2.349 & 1.090 & 0.619 & -0.354\\
PDC2 & 2.774 & 1.179 & 0.195 & -0.815 & 3.626 & 1.143 & -0.412 & -0.765 & 2.216 & 1.054 & 0.739 & -0.043\\
PDC3 & 2.666 & 1.176 & 0.281 & -0.758 & 3.684 & 1.150 & -0.492 & -0.716 & 2.236 & 1.123 & 0.796 & -0.100\\
RCM1 & 3.098 & 1.216 & -0.074 & -0.927 & 3.214 & 1.083 & -0.077 & -0.688 & 2.302 & 1.085 & 0.629 & -0.330\\
RCM2 & 3.078 & 1.181 & -0.064 & -0.893 & 3.167 & 1.131 & -0.061 & -0.777 & 2.326 & 1.099 & 0.686 & -0.164\\
RCM3 & 3.132 & 1.190 & -0.134 & -0.928 & 3.330 & 1.079 & -0.339 & -0.559 & 2.432 & 1.125 & 0.554 & -0.468\\
DSM1 & 4.135 & 1.081 & -1.232 & 0.766 & 4.439 & 0.875 & -1.762 & 3.102 & 4.023 & 1.078 & -0.984 & 0.151\\
DSM2 & 3.986 & 1.098 & -0.952 & 0.117 & 4.333 & 0.881 & -1.534 & 2.491 & 3.844 & 1.134 & -0.759 & -0.330\\
DSM3 & 3.963 & 1.112 & -0.827 & -0.213 & 4.235 & 0.922 & -1.156 & 0.960 & 3.635 & 1.157 & -0.456 & -0.808\\
\bottomrule
\end{tabular}}
\end{table*}

To check the reliability and factorial validity of the latent variables,
we first calculated several test indices (Cronbach's alpha, omega,
omega2, omega3 and average variance extracted; table 2). Cronbach's
alpha values are good, varying between .80 and .92, indicating a
satisfactory reliability of the latent variables. The average variance
extracted varies between min=.598 and max=.798, with values
\textgreater.50 regarded as good {[}49{]}.

In combination with considerable covariance among the latent factors
this raises questions concerning the discriminant validity of the latent
factors of the specified model {[}17{]}. Thus, a Fornell-Larcker test
was performed for each model, separately. For this test the squared
correlation between two factors is compared with the average variance
extracted for each factor. Here, the former needs to show a lower value
than the latter. As this was the case for all factors in the three
domains of application, the results suggest discriminant validity
between the latent factors within each group.

\begin{table*}

\caption{\label{tab:unnamed-chunk-8}Reliability Values}
\centering
\resizebox{\linewidth}{!}{
\fontsize{9}{11}\selectfont
\begin{tabular}[t]{lrrrrrrrrrrrrrrr}
\toprule
\multicolumn{1}{c}{ } & \multicolumn{5}{c}{Loan Origination} & \multicolumn{5}{c}{Job Recruitment} & \multicolumn{5}{c}{Medical Treatment} \\
\cmidrule(l{3pt}r{3pt}){2-6} \cmidrule(l{3pt}r{3pt}){7-11} \cmidrule(l{3pt}r{3pt}){12-16}
Item & RCG & PDC & RCM & DSM & Total & RCG & PDC & RCM & DSM & Total & RCG & PDC & RCM & DSM & Total\\
\midrule
alpha & 0.896 & 0.919 & 0.913 & 0.921 & 0.944 & 0.865 & 0.865 & 0.868 & 0.858 & 0.924 & 0.806 & 0.889 & 0.854 & 0.870 & 0.905\\
omega & 0.898 & 0.919 & 0.915 & 0.922 & 0.970 & 0.867 & 0.869 & 0.869 & 0.859 & 0.952 & 0.817 & 0.890 & 0.855 & 0.872 & 0.943\\
omega2 & 0.898 & 0.919 & 0.915 & 0.922 & 0.970 & 0.867 & 0.869 & 0.869 & 0.859 & 0.952 & 0.817 & 0.890 & 0.855 & 0.872 & 0.943\\
omega3 & 0.900 & 0.919 & 0.915 & 0.923 & 0.966 & 0.867 & 0.872 & 0.869 & 0.858 & 0.949 & 0.829 & 0.890 & 0.855 & 0.874 & 0.929\\
avevar & 0.745 & 0.790 & 0.781 & 0.798 & 0.777 & 0.685 & 0.689 & 0.689 & 0.671 & 0.685 & 0.598 & 0.729 & 0.662 & 0.696 & 0.673\\
\bottomrule
\end{tabular}}
\end{table*}

\hypertarget{measurement-invariance}{%
\subsection{Measurement Invariance}\label{measurement-invariance}}

Before addressing the hypotheses, we need to check for measurement
invariance, i. e. whether the measurement of the construct can be
considered invariant between groups. Only then a mean comparison of the
construct between groups is viable. A first CFA model addresses
configural invariance. In our model, the four latent factors are
measured with the three respective manifest indicators described in the
measurement section. Furthermore, we assume covariances between all
latent factors as every dimension reflects a special aspect of
perceptions regarding threats of AI. The chi-square test of model fit
reaches significance, \(\chi^2\)(144)=207.091, p\textless.001. In
addition, the approximate fit indices results show good fit for the
model, TLI=.989, RMSEA=.038 (.026, .050), SRMR=.026. Following the
suggestion by Vandenberg {[}57{]}, we do not automatically reject the present model
with high degrees of freedom and a considerable sample size on the basis
of the strict chi-square test, but look into the reasons for any
misspecification. Results show that the unexplained variance in the
specified model stems from cross-loadings of items RCG2 and RCG3 on the
dimension of prediction in the finance condition. While freeing the
respective parameters would improve model fit, at this point no such
action is taken.

In a first modification step, we calculated a CFA-model with measurement
invariance. Thus, we constraint the factor loadings of all items
assuming that the factors load equally on the latent factors in each
group. We compared the measurement invariance model with the original
model with configural invariance. The chi-square difference test shows
that the measurement invariance model does not fit the data worse than
the configural invariance model, \(\Delta\chi^2\)(16)=26.152, p=.052.
Accordingly, our data support the assumption of measurement invariance,
i. e. that the factor loadings are equal across different domains of
application.

In a second modification step, we calculated a CFA model with scalar
invariance. Accordingly, we constrained the intercepts of the items and
compared the fit of the model with the measurement invariance model. The
chi-square difference test suggests, that the scalar invariance model
performs significantly worse than the measurement invariance model,
\(\Delta\chi^2\)(16)=32.642, p=.008. To detect non-invariant intercepts
across the groups, we referred to the modification indices. These
suggest that the intercept of item DSM3 is non-invariant (modind =
7.44). Accordingly, we freed the intercept constraint of item DSM3 and
calculated a partial scalar invariance model. A chi-square difference
test for the model with partial scalar invariance fits the data not
worse than the measurement invariance model,
\(\Delta\chi^2\)(14)=21.311, p=.094.

In a third modification step, we also constraint the residuals and hence
calculated a model with strict invariance. According to our results, the
strict invariance model performs relatively poorly, \(\Delta\chi^2\)(24)
= 123.12, p\textless.001. Thus, the assumption of strict invariance is
rejected.

Consequently, the model with partial scalar invariance will be
discussed. The strict chi-square test for this model reaches
significance, \(\chi^2\)(174)=254.555, p\textless.001. Again, the
approximate fit indices show good fit for the model, TLI=.988,
RMSEA=.039 (.028, .050), SRMR=.037. Once more, allowing for
cross-loadings and correlated error terms of indicators from the same
latent factor would improve model fit. However, no action for
respecification is taken at this point.

Results suggest that the measurement in the first group from the domain of loan origination appears to be somewhat
problematic. Not only did the estimated model suggest that there are
unanticipated cross-loadings of indicators from the recognition function
to the latent factor of prediction, but also one item intercept of the
factor of decision-making is non-invariant. This serves as indication
that the measurement in the loan origination group did not work as
optimal as intended. However, model fit as indicated by the approximate
fit indices was still satisfactory.

Turning towards RQ1, we detect that individuals in fact have different
threat perceptions regarding distinct AI functionalities. Across all
tested domains respondents perceived recognition, prediction,
recommendation and decision-making as different, yet related,
functionalities of AI systems. This confirms our proposed measurement.
However, irrespective of the slightly problematic values in the loan
origination domain, the distinction between the functionalities proves
to be quite stable and consequently appears to be feasible.

\hypertarget{mean-differences-of-ai-functions-between-conditions}{%
\subsection{Mean differences of AI functions between
conditions}\label{mean-differences-of-ai-functions-between-conditions}}

After having established partial scalar invariance, the next step is to
address the mean comparisons between the three domains to test H1. H1
states that there are differences regarding the threat perception of
each function between the domains of AI application. The first domain
regarding the application of AI in loan origination serves as a
reference group. Accordingly, the means of the four latent factors of
the four functions in this group are constrained to zero.

Results show that compared with the domain 2 (`job recruitment')
prediction (\(\Delta\) M=.918, p\textless.001) and decision-making
(\(\Delta\) M=.327, p\textless.001) appeared to be significantly more
threatening in the job recruitment domain, while recognition (\(\Delta\)
M=.082, p=.355) and recommendation (\(\Delta\) M=.137, p\textless.116)
did not differ between both conditions. Thus, the job recruitment domain
was perceived as more threatening in two out of four functionalities
than the loan origination condition.

Compared with domain 3 (`medical treatment') recognition (\(\Delta\)
M=-1.190, p\textless.001), prediction (\(\Delta\) M=-.501,
p\textless.001) and recommendation (\(\Delta\) M=-.763, p\textless.001)
were perceived as significantly less threatening in the medical
treatment domain, while with regard to decision-making (\(\Delta\)
M=-.125, p=.139) there was no difference. Here, the job recruitment
domain was perceived as more threatening in three out of four
functionalities.

When comparing domain 2 (`job recruitment') with domain 3 (`medical
treatment') the results indicate that recognition (\(\Delta\)
M=-1.273, p\textless.001), prediction (\(\Delta\) M=-1.419,
p\textless.001), recommendation (\(\Delta\) M=-.900 , p\textless.001),
and decision-making (\(\Delta\) M=-.452, p\textless.001) all differed
significantly between both groups. In other words, the AI application
for the treatment of health problems was deemed less threatening than an
AI that was deployed to assess candidates for a job.

Summing up the results, the usage of AI in medical treatment is
perceived as less threatening in nearly all functionalities compared to
the usage of AI for loan origination or job recruitment. On the other
hand, the job recruitment domain was rated as most threatening compared
to the other domains - at least regarding the prediction and the
decision-making functionality.

Consequently, H1 that assumed differences in the threat perceptions
between the different domains of AI application was partially accepted.
Threat perceptions regarding AI functionalities appear to be widely
domain-dependent.

\hypertarget{effects-on-fear}{%
\subsection{Effects on Fear}\label{effects-on-fear}}

\begin{figure*}
\includegraphics[width=0.98\textwidth]{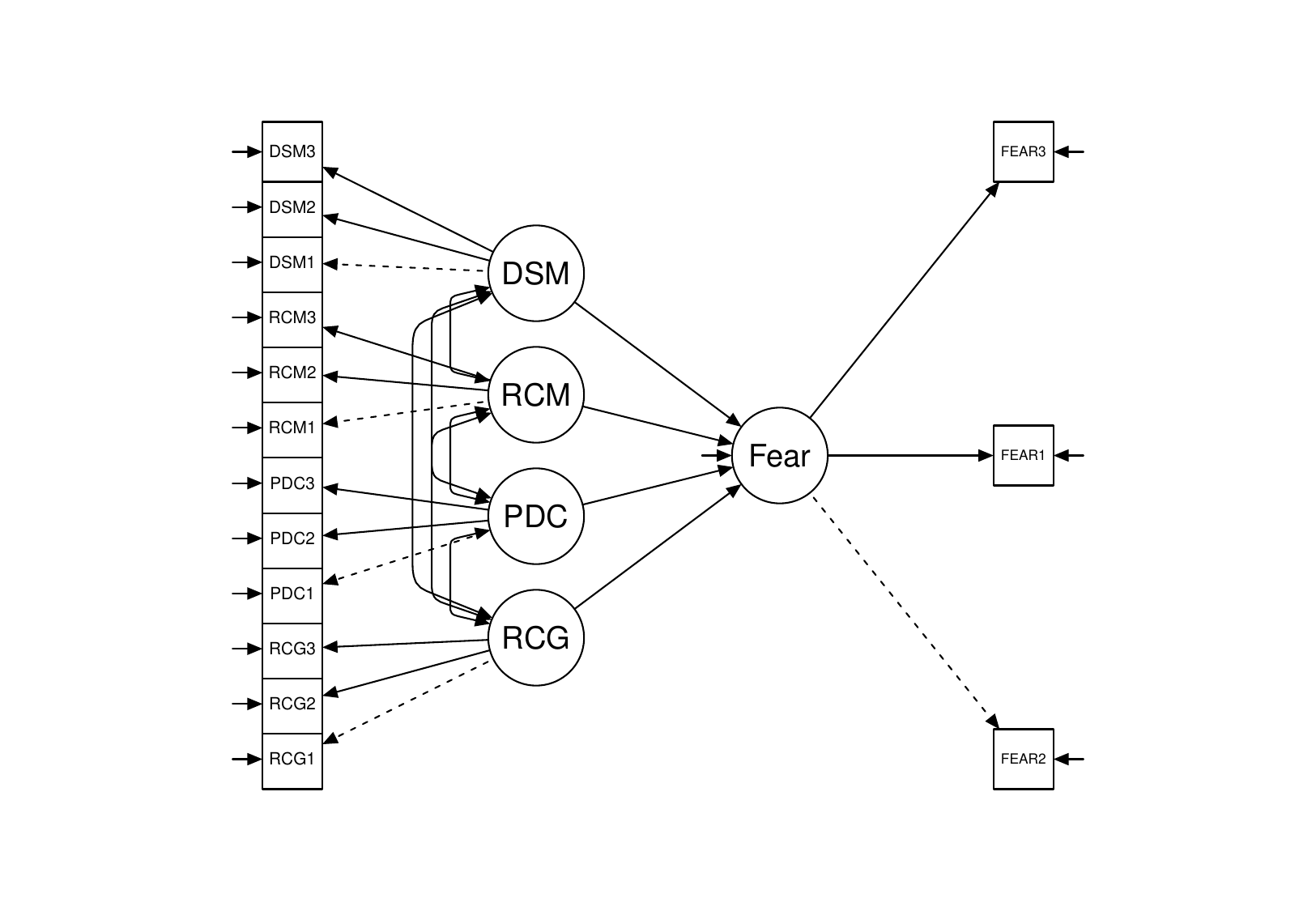} \caption{SEM Model}\label{fig:two-col-tribute-plot}
\end{figure*}

To assess whether threat perceptions are able to predict reported fear
of the respondents, a structural regression model was specified. Here
the three items that served as indicators of fear loaded on a latent
factor that is modeled as an endogenous dependent variable. The four
latent factors of threat perceptions of AI systems functions are modeled
as exogenous independent variables. Figure 1 displays the model.

Before addressing the analysis, measurement invariance of the fear
construct was assessed. The results suggest that measurement of the fear
construct is non-invariant between the three groups. Especially, the
loading of item FEAR1 is considerably non-invariant. Consequently, the
equality constraints for the loadings and intercept of this item were
freed for the analysis. The chi-square difference test indicates that
this model with partial invariance still fits significantly worse
compared to the configural model, \(\Delta\chi^2\)(4)=11.95, p=.018.
However, there is a substantial improvement in the comparative fit
indices between the model with measurement invariance and scalar
invariance, \(\Delta\)TLI=.003. While the actual measurement of the fear
construct is not optimal, we still rely on this previously tested
measurement of fear from the literature for the purpose of an analysis
of the connection between threat perceptions and self-reported fear.

Concerning the eventual analysis, in a first step, a model was estimated
where the regressions coefficient of threat perceptions on the induced
fear could vary freely, \(\chi^2\)(274)=376.578, p\textless.001. Again,
the approximate fit indices show good fit for the model, TLI=.987,
RMSEA=.036 (.026, .044), SRMR=.039. However, as the threat perceptions
are highly correlated this inflates the standard errors of the estimated
parameters and it needs to be tested whether their respective effects
actually differ between each other {[}20, 25{]}. Accordingly, due to
high inter-correlations of the four latent factors in the respective
groups, a second model was specified in which the effect of each latent
factor on fear was constrained to be equal. The chi-square difference
test for the first unconstrained model and the model with equality
constraints shows that the second model does not fit the data
significantly worse, \(\Delta\chi^2\)(9)=12.976, p=.164.

The model with equality constraints for the threat perceptions within
each group still suggests good fit, \(\chi^2\)(283)=389.555,
p\textless.001, TLI=.987, RMSEA=.036 (.026, .044), SRMR=.041. The
parameter estimates show a small effect of the threat perceptions of AI
on fear for group 1 (`loan origination'), B(SE)=.184(.014), p \textless.
001, group 2 (`job application'), B(SE)=.189(.018), p \textless. 001,
and group 3 (`medical treatment'), B(SE)=.188(.017), p \textless. 001,
respectively. The effect size ranges from \(\beta_{min}\)=.135 to
\(\beta_{max}\)=.203. Accordingly, H2 was accepted.

Eventually, RQ2 asks about differences of the effect of threat
perceptions on fear between the domains. The similarity of the parameter
estimates suggests that the effect of threat perceptions of AI on fear
appears to be equal between the groups. Consequently, a model was
specified where the effect of the latent factors of threat perceptions
on fear were not only equal across threat perceptions of the functions
of AI within the respective groups, but also did not differ between the
domains. Therefore, it was tested whether the effect of the threat
perceptions on fear was equal across the groups by using equality
constraints. The chi-square difference test for the first unconstrained
model and the model with equality constraints shows that the second
model does not fit the data significantly worse,
\(\Delta\chi^2\)(2)=0.059, p=.971.

The model with equality constraints for the effect of threat perceptions
of AI on perceived fear within and between each domain still suggests
good fit, \(\chi^2\)(285)=389.614, p\textless.001, TLI=.987, RMSEA=.035
(.026, .044), SRMR=.041. The parameter estimates show that threat
perceptions of AI have a small significant effect on reported fear for
group 1 (`loan origination'), group 2 (`job application'), and group 3
(`medical treatment'), respectively, B(SE)=.186(.010), p \textless. 001.
The effect size ranges from \(\beta_{min}\)=.133 to
\(\beta_{max}\)=.202.

\hypertarget{discussion}{%
\section{Discussion}\label{discussion}}

In this paper we introduced a scale to measure threat perceptions of
artificial intelligence. The scale can be used to assess citizens'
concerns regarding the use of AI systems. As AI technologies are
increasingly introduced into everyday life, it is crucial to understand
under what circumstances citizens might feel themselves to be
threatened. Threats can be understood as a pre-condition of fear.
Subsequently, according to fear appeal literature, being frightened can
lead to denial and avoidance of the threatening object. Thus, if people
perceive AI as a serious threat, it could cause a non-adoption of the
technology.

However, AI is an umbrella term for a huge variety of different
applications. AI applications can fulfill various functions and
applications are used in almost every societal field. Arguably, there
are huge variances in threat perceptions of different functions and
domains of application. With the TAI-scale we propose a measurement to
account for this context-specificity.

First, the results suggest that threat perceptions of distinct AI
functions can be reliably differentiated by respondents. Recognition,
prediction, recommendation and decision-making are indeed perceived as
different functions of AI systems. However, depending on the context
evaluated the measure showed diverging factorial validity. In one case
the indicator items had significant shared variance with more than one
dimension. This impairment of discriminatory power indicates that
thorough pre-testing of the adapted measures and data quality control
are of utmost importance when devising the survey instrument in
subsequent study designs. In doing so, researchers need to make sure
that respondents fully comprehend the item wording and that the object
of potential threat is clearly recognizable. Especially, this becomes
important when respondents are confronted with new and technically
sophisticated AI systems, for which there not yet exists enough direct
personal experience.

Second, threat perceptions are shown to vary between different domains,
in which AI systems are deployed. This suggests that the notion of a
general fear of AI needs to enhanced in favor of a broader conception
not only of what actions AI is able to perform, but also what exactly is
at stake in a given situation. In cases where AI systems seem useful and
the consequences of its application appear insubstantial, the
introduction of AI in another domain might evoke entirely opposite
reactions. Thus, while general perceptions such as general
predispositions concerning digital technology certainly do play a role
when it comes to the evaluation of innovative AI systems, a more
fine-grained approach is necessary and appears to be fruitful with the
developed measurement. Respondents' threat perceptions in this study
varied considerably between domains. Especially, the use of AI in
medical treatment was only perceived as lightly threatening, whereas
threat perceptions were quite higher in the domains of job recruitment
and loan origination. Regarding the levels of threat perceptions
concerning the functionalities, it is evident that the decision-making
function is perceived as most threatening within all three domains.
Arguably, this might be based on the loss of humans' autonomy. As this
is only a hypothesis at this point, further studies should elaborate on
these findings. Future applications of the TAI scale will yield further
insights concerning the items' and scales' sensitivity with regards to
different domains of AI applications.

Third, threat perceptions are reliable predictors of self-reported fear.
As the measurement of actual fear is rather complicated via the means of
a survey instrument, the inquiry of threat perceptions appears not only
to be preferable. It also suggests that it is reasonably well connected
to individual self-reports of experienced fear. Further studies should
elaborate on these findings and focus on the behavioral impact of
AI-related threat perceptions. As the fear appeal literature suggests,
one might expect that high levels of AI-induced fear lead to rejection
of the technology or even protest behavior.

\hypertarget{limitations}{%
\section{Limitations}\label{limitations}}

As this study attempts to develop a more fine-grained approach to
measuring threat perceptions regarding AI, its focus lay on scale
construction and testing the application of the scale in an online
survey. The results are thus limited to German online users from a
non-representative online access panel. Further research should extend
the scope of the domain of AI applications as well as addressing further
groups of stakeholders and, especially, behavioral consequences of
perceived threats of AI. Furthermore, a translation of the scale to
other languages appears as another promising avenue. As {[}19{]} showed,
based on longitudinal data from the Eurobarometer attitudes towards
robots and autonomous systems vary between countries and might be
subject to cultural influences that warrant research illuminating
divergent perceptions and their antecedents.

\hypertarget{conclusion}{%
\section{Conclusion}\label{conclusion}}

The public perception of Artificial Intelligence will become
increasingly important as applications that make use of AI technologies
will further proliferate in various societal domains. A populace that
perceives AI as threatening and that in consequence fears its
proliferation may prove as detrimental as a blind trust in the
benevolence of actors that implement AI systems as well as a general
overestimation of the veracity of assertions and decisions made by AI.
Consequently, the survey of threat perceptions of various AI systems is
of great research interest. In this paper, we proposed and constructed a
measurement of threat perceptions regarding AI that is able to capture
various functions performed by AI systems and that is adaptable to any
context of application that is of interest. The developed TAI scale
showed satisfactory results in that it reliably captured threat
perceptions regarding the distinct functions of recognition, prediction,
recommendation, and decision-making by AI. The results also suggest that
the developed scale is able to elucidate differences in these threat
perceptions between distinct domains of AI applications.

\hypertarget{references}{%
\section*{References}\label{references}}
\addcontentsline{toc}{section}{References}

\hypertarget{refs}{}
\leavevmode\hypertarget{ref-Acquisti.2015}{}%
{[}1{]} Acquisti, A. et al. 2015. Privacy and human behavior in the age
of information. \emph{Science (New York, N.Y.)}. 347, 6221 (2015),
509--514. 
\url{https://doi.org/10.1126/science.aaa1465}

\leavevmode\hypertarget{ref-Araujo.2020}{}%
{[}2{]} Araujo, T. et al. 2020. In AI we trust? Perceptions about
automated decision-making by artificial intelligence. \emph{AI \&
Society}. 21, 6 (2020).
\url{https://doi.org/10.1007/s00146-019-00931-w}

\leavevmode\hypertarget{ref-Bauer.1995}{}%
{[}3{]} Bauer, M.W. ed. 1995. \emph{Resistance to new technology:
Nuclear power, information technology, and biotechnology}. Cambridge
University Press.
\url{https://doi.org/10.1017/CBO9780511563706}

\leavevmode\hypertarget{ref-bitkom.2018}{}%
{[}4{]} bitkom 2018. Digitalisierung gestalten mit dem Periodensystem
der Künstlichen Intelligenz: Ein Navigationssystem für Entscheider. [\emph{Designing digitization with the periodic table
of Artificial Intelligence: A navigation system for decision makers.}]
Retrieved from
\url{https://www.bitkom.org/sites/default/files/2018-12/181204_LF_Periodensystem_online_0.pdf}

\leavevmode\hypertarget{ref-Bol.2018}{}%
{[}5{]} Bol, N. et al. 2018. Understanding the effects of
personalization as a privacy calculus: Analyzing self-disclosure across
health, news, and commerce contexts. \emph{Journal of Computer-Mediated
Communication}. 23, 6 (2018), 370--388.
\url{https://doi.org/10.1093/jcmc/zmy020}

\leavevmode\hypertarget{ref-Bourne.2019}{}%
{[}6{]} Bourne, C. 2019. AI cheerleaders: Public relations,
neoliberalism and artificial intelligence. \emph{Public Relations
Inquiry}. 8, 2 (2019), 109--125.
\url{https://doi.org/10.1177/2046147X19835250}

\leavevmode\hypertarget{ref-Brennen.2018}{}%
{[}7{]} Brennen, J.S. et al. 2018. An industry-led debate: How UK media
cover artificial intelligence. Reuters Institute for the Study of
Journalism.

\leavevmode\hypertarget{ref-Cave.2019}{}%
{[}8{]} Cave, S. et al. 2019. Scary robots: Examining public responses
to AI. \emph{Proceedings of the 2019 AAAI/ACM Conference on AI, Ethics,
and Society} (New York, NY, USA, 2019), 331--337.
\url{https://doi.org/10.1145/3306618.3314232}

\leavevmode\hypertarget{ref-Choi.2016}{}%
{[}9{]} Choi, E. et al. 2016. Doctor AI: Predicting clinical events via
recurrent neural networks. \emph{Proceedings of Machine Learning for
Healthcare}. 56, (2016).

\leavevmode\hypertarget{ref-Dehne.2017}{}%
{[}10{]} Dehne, M. 2017. \emph{Soziologie der Angst [Sociology of fear]}. Springer
Fachmedien Wiesbaden.
\url{https://doi.org/10.1007/978-3-658-15523-0}

\leavevmode\hypertarget{ref-Dilsizian.2014}{}%
{[}11{]} Dilsizian, S.E. and Siegel, E.L. 2014. Artificial intelligence
in medicine and cardiac imaging: Harnessing big data and advanced
computing to provide personalized medical diagnosis and treatment.
\emph{Current Cardiology Reports}. 16, 1 (2014), 441.
\url{https://doi.org/10.1007/s11886-013-0441-8}

\leavevmode\hypertarget{ref-Epskamp.2019}{}%
{[}12{]} Epskamp, S. et al. 2019. Package ``semPlot'': Path diagrams and
visual analysis of various sem packages' Output (v.1.1.2).
Retrieved from
\url{https://cran.r-project.org/web/packages/semPlot/semPlot.pdf}

\leavevmode\hypertarget{ref-Eubanks.2018}{}%
{[}13{]} Eubanks, V. 2018. \emph{Automating inequality: How high-tech
tools profile, police, and punish the poor}. St. Martin's Press.

\leavevmode\hypertarget{ref-EuropeanCommission.2017}{}%
{[}14{]} European Commission 2017. Special Eurobarometer 460. Attitudes
towards the impact of digitisation and automation on daily life.
\url{https://doi.org/10.2759/835661}

\leavevmode\hypertarget{ref-EuropeanCommission.2020}{}%
{[}15{]} European Commission 2020. Special Eurobarometer 496.
Expectations and concerns of connected and automated driving.

\leavevmode\hypertarget{ref-Fast.2017}{}%
{[}16{]} Fast, E. and Horvitz, E. 2017. Long-term trends in the public
perception of artificial intelligence. \emph{Thirty-First AAAI
Conference on Artificial Intelligence.} (2017).

\leavevmode\hypertarget{ref-Fornell.1981}{}%
{[}17{]} Fornell, C. and Larcker, D.F. 1981. Evaluating structural
equation models with unobservable variables and measurement error.
\emph{Journal of Marketing Research}. 18, 1 (1981), 39.
\url{https://doi.org/10.2307/3151312}

\leavevmode\hypertarget{ref-Ghodselahi.2011}{}%
{[}18{]} Ghodselahi, A. and Amirmadhi, A. 2011. Application of
artificial intelligence techniques for credit risk evaluation.
\emph{International Journal of Modeling and Optimization}. (2011),
243--249.
\url{https://doi.org/10.7763/IJMO.2011.V1.43}

\leavevmode\hypertarget{ref-Gnambs.2019}{}%
{[}19{]} Gnambs, T. and Appel, M. 2019. Are robots becoming unpopular?
Changes in attitudes towards autonomous robotic systems in europe.
\emph{Computers in Human Behavior}. 93, (2019), 53--61.
\url{https://doi.org/10.1016/j.chb.2018.11.045}

\leavevmode\hypertarget{ref-Grewal.2004}{}%
{[}20{]} Grewal, R. et al. 2004. Multicollinearity and measurement error
in structural equation models: Implications for theory testing.
\emph{Marketing Science}. 23, 4 (2004), 519--529.
\url{https://doi.org/10.1287/mksc.1040.0070}

\leavevmode\hypertarget{ref-Gurovich.2018}{}%
{[}21{]} Gurovich, Y. et al. 2018. DeepGestalt - identifying rare
genetic syndromes using deep learning.
\url{http://arxiv.org/pdf/1801.07637v1}

\leavevmode\hypertarget{ref-Hinks.2020}{}%
{[}22{]} Hinks, T. 2020. Fear of robots and life satisfaction.
\emph{International Journal of Social Robotics}. 98, 4 (2020), 792.
\url{https://doi.org/10.1007/s12369-020-00640-1}

\leavevmode\hypertarget{ref-Hmoud.2019}{}%
{[}23{]} Hmoud, B. and Varallyai, L. 2019. Will artificial intelligence
take over humanresources recruitment and selection? \emph{Network
Intelligence Studies}. 13, 7 (2019).

\leavevmode\hypertarget{ref-Hofmann.2020}{}%
{[}24{]} Hofmann, P. et al. 2020. Developing purposeful AI use cases - a
structured method and its application in project management. (2020).

\leavevmode\hypertarget{ref-Jagpal.1982}{}%
{[}25{]} Jagpal, H.S. 1982. Multicollinearity in structural equation
models with unobservable variables. \emph{Journal of Marketing
Research}. 19, 4 (1982), 431--439.
\url{https://doi.org/10.1177/002224378201900405}

\leavevmode\hypertarget{ref-Jiang.2017}{}%
{[}26{]} Jiang, F. et al. 2017. Artificial intelligence in healthcare:
Past, present and future. \emph{Stroke and vascular neurology}. 2, 4
(2017), 230--243.
\url{https://doi.org/10.1136/svn-2017-000101}

\leavevmode\hypertarget{ref-Jorgensen.2019}{}%
{[}27{]} Jorgensen, T.D. et al. 2019. SemTools: Useful tools for
structural equation modeling. R package version 0.5-2.
Retrieved from
\url{https://CRAN.R-project.org/package=semTools}

\leavevmode\hypertarget{ref-Katz.2017}{}%
{[}28{]} Katz, Y. 2017. Manufacturing an artificial intelligence
revolution. \emph{SSRN Electronic Journal}. (2017).
\url{https://doi.org/10.2139/ssrn.3078224}

\leavevmode\hypertarget{ref-Kelley.2019}{}%
{[}29{]} Kelley, P.G. et al. 2019. Happy and assured that life will be
easy 10 years from now: Perceptions of artificial intelligence in 8
countries.
\url{http://arxiv.org/pdf/2001.00081v1}

\leavevmode\hypertarget{ref-Laakasuo.2018}{}%
{[}30{]} Laakasuo, M. et al. 2018. What makes people approve or condemn
mind upload technology? Untangling the effects of sexual disgust, purity
and science fiction familiarity. \emph{Palgrave Communications}. 4, 1
(2018), 1--14.
\url{https://doi.org/10.1057/s41599-018-0124-6}

\leavevmode\hypertarget{ref-Leiner.2016}{}%
{[}31{]} Leiner, D.J. 2016. Our research's breadth lives on convenience
samples a case study of the online respondent pool ``SoSci Panel''.
\emph{Studies in Communication \textbar{} Media}. 5, 4 (2016), 367--396.
\url{https://doi.org/10.5771/2192-4007-2016-4-367}

\leavevmode\hypertarget{ref-Liang.2017}{}%
{[}32{]} Liang, Y. and Lee, S.A. 2017. Fear of autonomous robots and
artificial intelligence: Evidence from national representative data with
probability sampling. \emph{International Journal of Social Robotics}.
9, 3 (2017), 379--384.
\url{https://doi.org/10.1007/s12369-017-0401-3}

\leavevmode\hypertarget{ref-Maloney.2011}{}%
{[}33{]} Maloney, E.K. et al. 2011. Fear appeals and persuasion: A
review and update of the extended parallel process model. \emph{Social
and Personality Psychology Compass}. 5, 4 (2011), 206--219.
\url{https://doi.org/10.1111/j.1751-9004.2011.00341.x}

\leavevmode\hypertarget{ref-Marcinkowski.2020}{}%
{[}34{]} Marcinkowski, F. et al. 2020. Implications of ai (un-)fairness
in higher education admissions. \emph{Proceedings of the
ACM~FAT* Conference} (2020), 122--130.
\url{https://doi.org/10.1145/3351095.3372867}

\leavevmode\hypertarget{ref-Maurer.2016}{}%
{[}35{]} Maurer, M. et al. eds. 2016. \emph{Autonomous driving:
Technical, legal and social aspects}. Springer.

\leavevmode\hypertarget{ref-MayerSchonberger.2013}{}%
{[}36{]} Mayer-Schönberger, V. and Cukier, K. 2013. \emph{Big data: A
revolution that will transform how we live, work and think}. John
Murray.

\leavevmode\hypertarget{ref-McClure.2018}{}%
{[}37{]} McClure, P.K. 2018. ``You're fired,'' says the robot.
\emph{Social Science Computer Review}. 36, 2 (2018), 139--156.
\url{https://doi.org/10.1177/0894439317698637}

\leavevmode\hypertarget{ref-Mongeau.2012}{}%
{[}38{]} Mongeau, P.A. 2012. Fear appeals. \emph{The sage handbook of
persuasion: Developments in theory and practice}. J. Dillard and L.
Shen, eds. SAGE Publications, Inc. 184--199.
\url{https://doi.org/10.4135/9781452218410.n12}

\leavevmode\hypertarget{ref-Moors.2013}{}%
{[}39{]} Moors, A. et al. 2013. Appraisal theories of emotion: State of
the art and future development. \emph{Emotion Review}. 5, 2 (2013),
119--124.
\url{https://doi.org/10.1177/1754073912468165}

\leavevmode\hypertarget{ref-Nabi.2002}{}%
{[}40{]} Nabi, R.L. 2002. Discrete emotions and persuasion. \emph{The
persuasion handbook: Developments in theory and practice}. J.P. Dillard
and M. Pfau, eds. SAGE Publications, Inc. 289--308.

\leavevmode\hypertarget{ref-Ooms.2015}{}%
{[}41{]} Ooms, J. et al. 2015. The EPPM put to the test. \emph{Dutch
Journal of Applied Linguistics}. 4, 2 (2015), 241--256.
\url{https://doi.org/10.1075/dujal.4.2.07oom}

\leavevmode\hypertarget{ref-Pellegrino.2015}{}%
{[}42{]} Pellegrino, G. 2015. Obsolescence, presentification,
revolution: Sociotechnical discourse as site for in fieri futures.
\emph{Current Sociology}. 63, 2 (2015), 216--227.
\url{https://doi.org/10.1177/0011392114556584}

\leavevmode\hypertarget{ref-Raghavan.2020}{}%
{[}43{]} Raghavan, M. et al. 2020. Mitigating bias in algorithmic
hiring: Evaluating claims and practices. \emph{Proceedings of the
ACM~FAT* Conference} (2020), 469--481.
\url{https://doi.org/10.1145/3351095.3372828}

\leavevmode\hypertarget{ref-Renaud.2006}{}%
{[}44{]} Renaud, D. and Unz, D. 2006. Die M-DAS - eine modifizierte
Version der Differentiellen Affekt Skala zur Erfassung von Emotionen bei
der Mediennutzung. [\emph{The M-DAS - a modified version of the Differential Affect Scale for measuring emotions in media use.}]. \emph{Zeitschrift für Medienpsychologie}. 18, 2
(2006), 70--75.
\url{https://doi.org/10.1026/1617-6383.18.2.70}

\leavevmode\hypertarget{ref-RodriguezRuiz.2019}{}%
{[}45{]} Rodriguez-Ruiz, A. et al. 2019. Stand-alone artificial
intelligence for breast cancer detection in mammography: Comparison with
101 radiologists. \emph{Journal of the National Cancer Institute}. 111,
9 (2019), 916--922.
\url{https://doi.org/10.1093/jnci/djy222}

\leavevmode\hypertarget{ref-Rogers.1975}{}%
{[}46{]} Rogers, R.W. 1975. A protection motivation theory of fear
appeals and attitude change. \emph{The Journal of Psychology}. 91, 1
(1975), 93--114.
 \url{https://doi.org/10.1080/00223980.1975.9915803}

\leavevmode\hypertarget{ref-Rosseel.2012}{}%
{[}47{]} Rosseel, Y. 2012. Lavaan : An r package for structural equation
modeling. \emph{Journal of Statistical Software}. 48, 2 (2012).
\url{https://doi.org/10.18637/jss.v048.i02}

\leavevmode\hypertarget{ref-SanchezMonedero.2020}{}%
{[}48{]} Sánchez-Monedero, J. et al. 2020. What does it mean to 'solve'
the problem of discrimination in hiring? \emph{Proceedings of the
ACM~FAT* Conference} (2020), 458--468.
\url{https://doi.org/10.1145/3351095.3372849}

\leavevmode\hypertarget{ref-Segars.1997}{}%
{[}49{]} Segars, A.H. 1997. Assessing the unidimensionality of
measurement: A paradigm and illustration within the context of
information systems research. \emph{Omega}. 25, 1 (1997), 107--121.
\url{https://doi.org/10.1016/S0305--0483(96)00051--5}

\leavevmode\hypertarget{ref-Shen.2017}{}%
{[}50{]} Shen, L. 2017. Putting the fear back again (and within
individuals): Revisiting the role of fear in persuasion. \emph{Health
Communication}. 32, 11 (2017), 1331--1341.
\url{https://doi.org/10.1080/10410236.2016.1220043}

\leavevmode\hypertarget{ref-Smith.1990}{}%
{[}51{]} Smith, C.A. and Lazarus, R.S. 1990. Emotion and adaptation.
\emph{Handbook of personality : Theory and research}. L.A. Pervin, ed.
Guilford Pr. 609--637.

\leavevmode\hypertarget{ref-So.2013}{}%
{[}52{]} So, J. 2013. A further extension of the extended parallel
process model (E-EPPM): Implications of cognitive appraisal theory of
emotion and dispositional coping style. \emph{Health Communication}. 28,
1 (2013), 72--83.
 \url{https://doi.org/10.1080/10410236.2012.708633}

\leavevmode\hypertarget{ref-So.2016}{}%
{[}53{]} So, J. et al. 2016. Reexamining fear appeal models from
cognitive appraisal theory and functional emotion theory perspectives.
\emph{Communication Monographs}. 83, 1 (2016), 120--144.
\url{https://doi.org/10.1080/03637751.2015.1044257}

\leavevmode\hypertarget{ref-Tambe.2019}{}%
{[}54{]} Tambe, P. et al. 2019. Artificial intelligence in human
resources management: Challenges and a path forward. \emph{California
Management Review}. 61, 4 (2019), 15--42.
\url{https://doi.org/10.1177/0008125619867910}

\leavevmode\hypertarget{ref-Topol.2019}{}%
{[}55{]} Topol, E.J. 2019. High-performance medicine: The convergence of
human and artificial intelligence. \emph{Nature Medicine}. 25, 1 (2019),
44--56.
\url{https://doi.org/10.1038/s41591-018-0300-7}

\leavevmode\hypertarget{ref-Tudor.2003}{}%
{[}56{]} Tudor, A. 2003. A (macro) sociology of fear? \emph{The
Sociological Review}. 51, 2 (2003), 238--256.
\url{https://doi.org/10.1111/1467-954X.00417}

\leavevmode\hypertarget{ref-Vandenberg.2006}{}%
{[}57{]} Vandenberg, R.J. 2006. Introduction: Statistical and
methodological myths and urban legends. \emph{Organizational Research
Methods}. 9, 2 (2006), 194--201.
\url{https://doi.org/10.1177/1094428105285506}

\leavevmode\hypertarget{ref-Wirtz.2019}{}%
{[}58{]} Wirtz, B.W. et al. 2019. Artificial intelligence and the public
sector---applications and challenges. \emph{International Journal of
Public Administration}. 42, 7 (2019), 596--615.
\url{https://doi.org/10.1080/01900692.2018.1498103}

\leavevmode\hypertarget{ref-Witte.1992}{}%
{[}59{]} Witte, K. 1992. Putting the fear back into fear appeals: The
extended parallel process model. \emph{Communication Monographs}. 59, 4
(1992), 329--349.
\url{https://doi.org/10.1080/03637759209376276}

\leavevmode\hypertarget{ref-Zhang.2019}{}%
{[}60{]} Zhang, B. and Dafoe, A. 2019. Artificial intelligence: American
attitudes and trends. \emph{SSRN Electronic Journal}. (2019).
\url{https://doi.org/10.2139/ssrn.3312874}

\end{document}